\documentclass[letterpaper,10pt]{article}

\usepackage[utf8]{inputenc}
\usepackage{amsmath}
\usepackage{amssymb}

\title{BTZ quasinormal frequencies as poles of Green's function}

\author{A.\ L\'opez-Ortega\thanks{alopezo@ipn.mx}, D.\ Mata-Pacheco \\ 
Departamento de F\'{\i}sica. \\
Escuela Superior de F\'{\i}sica y Matem\'aticas. \\
Instituto Polit\'ecnico Nacional. \\
Unidad Profesional Adolfo L\'opez Mateos. Edificio 9. \\
Ciudad de M\'exico, M\'exico. \\
C.\ P.\ 07738 }

\begin{document}

\maketitle

\begin{abstract}

Based on the well known fact that the quasinormal frequencies are the poles of the frequency domain Green's function we describe a method that allows us to calculate exactly the  quasinormal frequencies of the Klein-Gordon field moving in the three-dimensional rotating BTZ  black hole. These quasinormal frequencies are already published and widely explored in several applications, but we use this example to expound the proposed method of computation. We think that the described procedure can be useful to calculate exactly the quasinormal frequencies of classical fields propagating in other backgrounds. Furthermore, we compare with previous results and discuss some related facts.



\end{abstract}

\section{Introduction}
\label{s: Introduction}

Several dynamical aspects of the  black hole perturbations are encoded in their quasinormal modes (QNM), namely in the solutions to their equations of motion that are purely ingoing near the event horizon. Also, in the far region they satisfy a boundary condition determined by the asymptotic structure of the spacetime, for example, for asymptotically flat black holes, it is imposed that the perturbation is purely outgoing in the far region, whereas for asymptotically anti-de Sitter black holes a Dirichlet type boundary condition is commonly imposed in the far region \cite{Kokkotas:1999bd}, \cite{Berti:2009kk}. Owing to these boundary conditions, the QNM possess complex frequencies, the quasinormal frequencies (QNF). As is well known, the real part of the QNF is the oscillation frequency of the perturbation and its imaginary part determines the decay time  \cite{Kokkotas:1999bd}, \cite{Berti:2009kk}. 

Motivated by the AdS-CFT correspondence, the QNF spectrum of asymptotically anti-de Sitter spacetimes has been studied,  because they determine the decay times of the dual quantum field theories living at the boundary \cite{Horowitz:1999jd}, \cite{Birmingham:2001pj}. As a consequence, the QNF spectrum of asymptotically anti-de Sitter black holes is widely explored, \cite{Kokkotas:1999bd}, \cite{Berti:2009kk}, and we find  numerical computations in many references, see for example Refs.\ \cite{Horowitz:1999jd}, \cite{Cardoso:2001bb}--\cite{Miranda:2008vb}, whereas in Refs.\ \cite{Birmingham:2001pj}, \cite{Cardoso:2001hn}--\cite{Oliva:2010xn}, we encounter exact calculations of the QNF for several test fields.

Among the asymptotically anti-de Sitter solutions, the three-dimensional Ba\~nados-Teitelboim-Zanelli black hole  (BTZ black hole in what follows) \cite{Banados:1992wn}, \cite{Banados:1992gq} plays a special role, since in this background we can obtain several exact results \cite{Birmingham:2001dt}. An example is the exact determination of the QNF spectra for the Klein-Gordon, Dirac, and massive vector fields propagating on it \cite{Birmingham:2001pj}, \cite{Cardoso:2001hn}, \cite{Birmingham:2001hc}. Here we use the widely known example of exact calculation for the QNF of the Klein-Gordon field moving in the BTZ black hole to describe an alternative method to compute exactly the proper oscillations of classical fields propagating in black holes. The proposed method is based on the work by Leaver of Ref.\ \cite{Leaver:1986gd} that defines the QNF as the poles of the frequency domain Green's function. As far as we know this method is not previously exploited  to compute exactly the QNF of black holes and in what follows we describe this procedure.

\section{QNF of the BTZ black hole}
\label{s: QNF of the BTZ black hole}

As is well known, in the coordinates $(t,r,\phi)$ the metric of the three-dimensional rotating BTZ black hole is\footnote{As in Ref.\ \cite{Birmingham:2001hc}, we choose units such that $8 G =1$.} \cite{Banados:1992wn}, \cite{Banados:1992gq}
\begin{equation} \label{e: metric BTZ} 
 ds^2 = - f(r) dt^2 +  f(r)^{-1} dr^2 + r^2 \left(d \phi -\frac{J}{2 r^2} dt\right)^2 ,
\end{equation} 
where the function $f$ is equal to
\begin{equation}
 f(r) = -M + \frac{r^2}{l^2} + \frac{J^2}{4r^2},
\end{equation} 
the factor $l^2$ is related to the negative cosmological constant by $\Lambda = -1/l^2$ and we point out that the mass $M$ and the angular momentum $J$ of the black hole are determined by
\begin{equation}
 M = \frac{r_+^2 + r_-^2}{l^2}, \qquad \qquad \qquad J = \frac{2 r_+ r_-}{l},
\end{equation} 
with $r_{\pm}$ denoting the outer and inner radii of the black hole \cite{Banados:1992wn}, \cite{Banados:1992gq}.

In analogy to Refs.\ \cite{Cardoso:2001hn}, \cite{Birmingham:2001hc}, we define the QNM of the BTZ black hole (\ref{e: metric BTZ}) as the modes that satisfy the boundary conditions:

a) The field is ingoing near the outer horizon.

b) The field goes to zero as $r \to \infty$.

For the cases in which an exact solution of the radial equation is available, the usual method for computing exactly the QNF is to impose the boundary conditions on the radial function and then we calculate exactly the QNF from the conditions on the radial solutions \cite{Cardoso:2001hn}--\cite{Oliva:2010xn}. Nevertheless a different method based on the results of Ref.\ \cite{Leaver:1986gd} is possible and in what follows we describe this procedure. Since the work by Leaver of Ref.\ \cite{Leaver:1986gd}, it is known that the QNF correspond to the poles of Green's function in the frequency domain (see also Refs.\ \cite{Kokkotas:1999bd}--\cite{Berti:2009kk}). Based on this picture and on the results by Birmingham \cite{Birmingham:2001hc} for the QNM of the Klein-Gordon field propagating in the  three-dimensional rotating BTZ black hole \cite{Banados:1992wn}, \cite{Banados:1992gq} we calculate exactly their QNF and compare with the previous results of Ref.\  \cite{Birmingham:2001hc}.\footnote{For the exact calculation of the QNF spectrum for the Klein-Gordon field moving in the static BTZ black hole see Ref.\  \cite{Cardoso:2001hn}.} As far as we know this method is not used previously in the exact computation of the QNF for asymptotically anti-de Sitter black holes \cite{Kokkotas:1999bd}, \cite{Berti:2009kk}, \cite{Cardoso:2001hn}--\cite{Oliva:2010xn}.

For the Schr\"odinger type equation \cite{Kokkotas:1999bd}, \cite{Berti:2009kk}
\begin{equation} \label{e: Schrodinger type}
 \frac{d^2 \Phi}{dx^2} + \omega^2 \Phi -V \Phi = 0, 
\end{equation} 
where $\Phi$ is related to the radial function of the field, $\omega$ denotes the frequency and $V$ is the effective potential that is characteristic of the field under study, it is well known that its frequency domain Green's function is equal to \cite{Leaver:1986gd}
\begin{equation}\label{e: Green definition}
\tilde{G}(x_{1},x_{2},\omega) = \begin{cases}
\frac{\Phi_{+}(\omega,x_{1}) \Phi_{\infty}(\omega,x_{2})}{W_x(\omega)}, & x_{1}<x_{2};\\
\frac{\Phi_{+}(\omega,x_{2}) \Phi_{\infty}(\omega,x_{1})}{W_x(\omega)},& x_{2}<x_{1} ,
\end{cases} 
\end{equation} 
where $\Phi_{+}$ is the  solution of the Schr\"odinger type equation satisfying the boundary condition near the horizon of the black hole. In  a similar way $\Phi_{\infty}$ fulfills the boundary condition as $r \to \infty$, and 
\begin{equation}\label{e: Wronskian definition}
W_x(\Phi_{\infty},\Phi_{+})  = W_x(\omega) = \Phi_{\infty}\dfrac{d \Phi_{+}}{d x}-\Phi_{+}\dfrac{d \Phi_{\infty}}{d x} , 
\end{equation}
is the Wronskian of these solutions. In the rest of this section, for the Klein-Gordon field we determine the Wronskian (\ref{e: Wronskian definition}) of the solutions satisfying the boundary conditions at the event horizon and at the asymptotic region. Taking as a basis the expression (\ref{e: Green definition}) for the frequency domain Green's function we calculate exactly the corresponding QNF by finding the zeros of its Wronskian, that is, the poles of the frequency domain Green's function.

The Klein-Gordon equation takes the form\footnote{In the rest of this work we follow closely the notation of Ref.\ \cite{Birmingham:2001hc}, but we make some small changes. }
\begin{equation}\label{Klein-Gordon equation}
\left( \Box^{2}-\frac{\mu^{2}}{l^2} \right) \Psi = \left( \frac{1}{\sqrt{|g|}}\partial_\mu (\sqrt{|g|} g^{\mu \nu} \partial_\nu) - \frac{\mu^{2}}{l^2} \right) \Psi =  0 , 
\end{equation}
where $\Box^{2}$ is the d'Alembertian operator, $\mu$ is related to the mass of the field, $g^{\mu \nu}$ denotes the inverse metric, $g$ its determinant  and $\Psi$ denotes the Klein-Gordon field. As shown in Ref.\  \cite{Birmingham:2001hc} if we take a separable solution
\begin{equation}
 \Psi = R(r) e^{-i \omega t} e^{i m \phi},
\end{equation}
where $m$ is the angular eigenvalue, then the Klein-Gordon equation simplifies to the following  differential equation for the radial function $R$
\begin{equation} \label{e: radial original}
\frac{d}{dr}\left( f \frac{d R}{dr} \right) + \frac{1}{r} f \frac{d R}{dr} + \left( \frac{\omega^2}{f} - \frac{J \omega m}{f r^2} - \frac{(r^2/l^2 - M)m^2}{f r^2}    \right) R = 0.
\end{equation} 
We notice that the previous expression can be transformed to a Schr\"odinger type equation of the form (\ref{e: Schrodinger type}) as follows 
\begin{equation} \label{e: Schrodinger type BTZ}
 \frac{d^2 \hat{R}}{dx^2} + \omega^2 \hat{R}  - \left\{ \frac{J \omega m}{r^2} + \frac{(r^2/l^2 - M)m^2}{r^2}  - \frac{f}{g}\frac{d}{dr}\left( f \frac{dg}{dr} \right)  - \frac{f}{g} \frac{dg}{dr} \right\} \hat{R} = 0,
\end{equation} 
with
\begin{equation} \label{e: relations operator radail}
 f  \frac{d}{dr} =  \frac{d}{dx} , \qquad g=   \frac{1}{\sqrt{r}} , \qquad R = g \hat{R} .
\end{equation} 
In Eq.\  (\ref{e: Schrodinger type BTZ}) the expression in curly braces is the effective potential of the Klein-Gordon field in the rotating BTZ black hole.

In the variable $z$ defined by 
\begin{equation}
 z = \frac{r^2-r_+^2}{r^2-r_-^2} ,
\end{equation} 
we find that the radial equation (\ref{e: radial original}) simplifies to \cite{Birmingham:2001hc}
\begin{equation}
 z (1-z) \frac{d^2 R}{dz^2}  + (1-z) \frac{d R}{ d z} + \left( \frac{A}{z} + B + \frac{C}{1-z} \right) R = 0 ,
\end{equation} 
where
\begin{align}
 A&=\frac{l^4 }{4 (r_+^2-r_-^2)^2} \left( \omega r_+ - \frac{m}{l}r_- \right)^2, \qquad C = - \frac{\mu^2}{4} , \nonumber \\
 B&=-\frac{l^4 }{4 (r_+^2-r_-^2)^2} \left( \omega r_- - \frac{m}{l}r_+ \right)^2.
\end{align}
Taking the radial function $R$ in the form 
\begin{equation} \label{e: ansatz radial}
 R= z^\alpha (1-z)^\beta F ,
\end{equation}  
we obtain that the function $F$ must be a solution to the hypergeometric differential equation \cite{Guo-book}--\cite{NIST-book}
\begin{equation}\label{e: hypergeometric equation}
z(1-z)\frac{d^{2} F }{dz^{2}} +\left[c-(a+b+1) z \right]\frac{d F }{dz} - a  b  F=0 ,
\end{equation}
when the quantities $\alpha$ and $\beta$ are solutions to the algebraic equations
\begin{equation}
 \alpha^2 + A = 0, \qquad \qquad \beta^2 - \beta + C= 0,
\end{equation} 
and the parameters of the hypergeometric equation are equal to
\begin{equation}\label{e: a b c hypergeometric}
 a = \alpha + \beta + i \sqrt{-B}, \qquad b = \alpha + \beta - i \sqrt{-B}, \qquad c = 2 \alpha + 1.
\end{equation}
As in Ref.\  \cite{Birmingham:2001hc}, in what follows we choose $\alpha = -i \sqrt{A}$ and $\beta = (1 -\sqrt{1+\mu^2})/2$.

Hence, around $z=0$ the solutions of the radial equation take the form
\begin{equation}
 R = z^\alpha (1-z)^\beta (C_1 \,_2F_1(a,b;c;z) + C_2 z^{1-c} \, _2F_1(a-c +1,b-c +1;2-c;z) ) ,
\end{equation} 
where $C_1$,  $C_2$ are constants and ${}_2F_1(a,b;c;z)$ denotes the hypergeometric function \cite{Guo-book}--\cite{NIST-book}. In a straightforward way we can verify that the solution satisfying the boundary condition a) near the outer horizon is \cite{Birmingham:2001hc}
\begin{equation}
 R_+ =  C_1 z^\alpha (1-z)^\beta \,_2F_1(a,b;c;z).
\end{equation} 

Instead of following the usual way to calculate exactly the QNF \cite{Birmingham:2001hc}, here we determine the solutions to the radial equation (\ref{e: radial original}) as $r \to \infty$, that is, around $z=1$. In what follows, it is convenient to utilize the variable $u$ defined by
\begin{equation} \label{e: u definition}
 u = 1-z .
\end{equation} 
Using this variable we find that the function $F$ of the formula (\ref{e: ansatz radial}) must be solution of a hypergeometric differential equation (\ref{e: hypergeometric equation}) with parameters \cite{Guo-book}
\begin{equation}
 \tilde{a} = a, \qquad 
 \tilde{b} = b , \qquad 
 \tilde{c} = 1 + a + b - c ,
\end{equation}
and therefore the radial function takes the form
\begin{align}
 R = (1-u)^\alpha u^\beta &(  K_1 {}_2F_1(a,b;a+b-c+1;u) \nonumber \\
 &+ K_2 u^{c-a-b} {}_2F_1(c-a,c-b;c-a-b+1;u) ),
\end{align} 
where $K_1$ and $K_2$ are constants. Analyzing its behavior as $r \to \infty$ we obtain that the radial function fulfilling the boundary condition b) at the asymptotic region is
\begin{equation}
 R_\infty = K_2 (1-u)^\alpha u^\beta u^{c-a-b} {}_2F_1(c-a,c-b;c-a-b+1;u),
\end{equation} 
since the term proportional to $K_1$ diverges as $u \to 0$.

Considering Eq.\ (\ref{e: relations operator radail}) we obtain that for the BTZ black hole $\Phi_+ = \sqrt{r} R_+$ and $\Phi_\infty = \sqrt{r} R_\infty$ and hence we get that their Wronskian is equal to 
\begin{equation}
 W_x (\Phi_{\infty},\Phi_{+})= r W_x (R_{\infty},R_{+}).
\end{equation} 
Owing to the hypergeometric differential equation (\ref{e: hypergeometric equation}) is of second order, we know that among any three solutions there is a linear relation \cite{Guo-book}--\cite{NIST-book}. An example of this type is the relation among its two solutions around $z=1$ and one of its solutions around $z=0$, usually known as Kummer property that takes the form 
\begin{align}\label{e: Identity Kummer}
&_{2}F_{1}(a,b; c;z)=\frac{\Gamma(c)\Gamma(c-a-b)}{\Gamma(c-a)\Gamma(c-b)}{}_{2}F_{1}(a,b;a+b-c+1;1-z) \\
&+ \frac{\Gamma(c)\Gamma(a+b-c)}{\Gamma(a)\Gamma(b)}(1-z)^{c-a-b}{}_{2}F_{1}(c-a,c-b;c-a-b+1;1-z) . \nonumber 
\end{align}

From the Kummer property of the hypergeometric function  and considering that  the Wronskian of the solutions to the hypergeometric equation is \cite{NIST-book}
\begin{align} \label{e: Wronskian hypergeometric}
&W_{z} \left(  \,_2F_1(a,b;a+b-c+1;1-z), (1-z)^{c-a-b} \right.  & \\ 
& \left. \times  \,_2F_1(c-a,c-b;c-a-b+1;1-z) \right) =(a+b-c)z^{-c}(1-z)^{c-a-b-1}  \nonumber , 
\end{align}
we find that the Wronskian of the radial solutions $\Phi_{+}$ and $\Phi_{\infty}$ is equal to 
\begin{equation}\label{e: Wronskian KG}
W_x (\Phi_{\infty},\Phi_{+})= -\frac{2}{l^2} K_2 C_{1} (r_+^2-r_-^2) (a+b-c) \dfrac{\Gamma(c)\Gamma(c-a-b)}{\Gamma(c-a)\Gamma(c-b)} .
\end{equation}
We notice that $W_x(\Phi_{\infty},\Phi_{+})$ does not depend on the coordinate $x$ (or $r$), as we expect \cite{Leaver:1986gd}. From the last expression we find that the zeros of the Wronskian (that is, the poles of Green's function) are located at \cite{Abramowitz-book}
\begin{equation}
 c-a=-n, \qquad \qquad c-b=-n, \qquad \qquad n=0,1,2,3,\ldots,
\end{equation} 
and therefore the QNF of the Klein-Gordon field in the rotating BTZ black hole are equal to  \cite{Birmingham:2001hc}
\begin{align}
 \omega & = \frac{m}{l} - \frac{2 i}{l^2} (r_+ - r_-) \left( n + \frac{1}{2} + \frac{\sqrt{1 + \mu^2}}{2} \right), \nonumber \\
 \omega & = -\frac{m}{l} - \frac{2 i}{l^2} (r_+ + r_-) \left( n + \frac{1}{2} + \frac{\sqrt{1 + \mu^2}}{2} \right),
\end{align}
that is, the proposed method produces the same QNF that the procedure used in Ref.\  \cite{Birmingham:2001hc}.

\section{Conclusions}
\label{s: Discussion}

The QNF of the rotating BTZ black hole have been useful in several applications \cite{Birmingham:2001pj}, \cite{Cardoso:2001hn}, \cite{Birmingham:2001hc}, \cite{Birmingham:2001dt}, and this is an example showing the utility of the exact results in the research line of black hole perturbations and related areas. In this paper our main objective is the description of the method given in the previous section and based on the results of Ref.\  \cite{Leaver:1986gd}. We believe that the proposed method can be used to calculate exactly the QNF of classical fields propagating in other black holes and this procedure is an useful addition to the tools for computing exactly the QNF of spacetimes, because we think that the possibility of calculating the physical quantities with different methods contributes to a better understanding of their properties. Furthermore, notice that using the proposed method  we can verify the exact results that are obtained with the usual procedure of Refs.\ \cite{Birmingham:2001pj}, \cite{Cardoso:2001hn}--\cite{Oliva:2010xn}. 

It is convenient to mention that for the Klein-Gordon field propagating in the rotating BTZ black hole, the proposed procedure yields the same values for the QNF as the usual method of Refs.\ \cite{Cardoso:2001hn}\--\cite{Oliva:2010xn}, but the developed procedure makes explicit the interpretation of the QNF as the poles of the frequency domain Green's function and this fact is not manifest in the procedure of Refs.\ \cite{Cardoso:2001hn}\--\cite{Oliva:2010xn}. We believe that this characteristic is a relevant addition of the method previously described in Sect. \ref{s: QNF of the BTZ black hole} and can be useful in the AdS-CFT correspondence \cite{Horowitz:1999jd}, \cite{Birmingham:2001pj}. Another feature of the proposed method is that for the solution satisfying the boundary condition near the horizon and for the solution fulfilling boundary condition at the asymptotic region, the procedure makes clear that at the QNF their Wronskian is equal to zero. A fact that is commonly stated \cite{Kokkotas:1999bd}, \cite{Berti:2009kk}, but not explicitly verified.

As far as we are aware this method is not previously used in the exact determination of the QNF for black holes, and we believe that this procedure can be an useful tool in the study of spacetime perturbations. Doubtless the applicability of the method that we expound in the present work to the determination of the QNM for other gravitational systems deserves further research.

\section{Acknowledgement} 

This work was supported by CONACYT M\'exico, SNI M\'exico, COFAA-IPN, EDI-IPN, and the Research Project IPN SIP-20181408.

\end{document}